# A new approach to local hardness

T. Gál[a,*], P. Geerlings,[a] F. De Proft[a], and M. Torrent-Sucarrat[b,*]

[a]Department of General Chemistry (Member of the QCMM Alliance Ghent-Brussels),
Free University of Brussels (VUB), Pleinlaan 2, 1050 Brussel, Belgium
[b]Departament de Química Biològica i Modelització Molecular,
IQAC–CSIC, c/ Jordi Girona 18, 08034 Barcelona, Spain

**Abstract:** The applicability of the local hardness as defined by the derivative of the chemical potential with respect to the electron density is undermined by an essential ambiguity arising from this definition. Further, the local quantity defined in this way does not integrate to the (global) hardness – in contrast with the local softness, which integrates to the softness. It has also been shown recently that with the conventional formulae, the largest values of local hardness do not necessarily correspond to the hardest regions of a molecule. Here, in an attempt to fix these drawbacks, we propose a new approach to define and evaluate the local hardness. We define a local chemical potential, utilizing the fact that the chemical potential emerges as the additive constant term in the number-conserving functional derivative of the energy density functional. Then, differentiation of this local chemical potential with respect to the number of electrons leads to a local hardness that integrates to the hardness, and possesses a favourable property; namely, within any given electron system, it is in a local inverse relation with the Fukui function, which is known to be a proper indicator of local softness in the case of soft systems. Numerical tests for a few selected molecules and a detailed analysis, comparing the new definition of local hardness with the previous ones, show promising results.

*Corresponding authors, e-mail: galt@phys.unideb.hu, mtsqbm@iqac.csic.es

## I. Introduction

Conceptual density functional theory (DFT)[1,2] offers a wide range of chemical reactivity indices for the description of chemical processes. Several of these indices have found successful application in the study of chemical phenomena. The three most well known reactivity descriptors, the electronegativity,[3] the chemical hardness, and softness,[4] have a long history, going back to times before the birth of DFT. They are basic constituents of essential principles governing chemical reactions – the electronegativity equalization principle,[5] the hard/soft acid/base principle,[4] and the maximum hardness principle.[6] An important aim of conceptual DFT is to establish local versions of the global indices on the basis of which predictions can be made regarding the molecular sites a given reaction happens. This would also make the establishment of local generalizations of the principles based on global reactivity descriptors possible. Pearson[4] himself already introduced a local version of the hard/soft acid/base principle, which made the exact foundation of a local hardness and a local softness index certainly desirable.

In the case of softness, defined as

$$S = \left(\frac{\partial N}{\partial \mu}\right)_{v(r)}, \quad (1)$$

a corresponding local quantity can be readily introduced; namely,

$$s(r) = \left(\frac{\partial n(r)}{\partial \mu}\right)_{v(r)}, \quad (2)$$

which is called the local softness,[7] and has a direct connection with the Fukui function $f(\vec{r}) = \left(\frac{\partial n(\vec{r})}{\partial N}\right)_{v(\vec{r})}$,[8] a well-established chemical reactivity index. Applying the chain rule of differentiation, one obtains

$$s(r) = \left(\frac{\partial n(r)}{\partial N}\right)_{v(r)} \left(\frac{\partial N}{\partial \mu}\right)_{v(r)} = f(r) S. \quad (3)$$

As it is easily seen, $s(r)$ integrates to $S$ (as the Fukui function integrates to 1), so it is natural to interpret it as a pointwise, *i.e.* local, softness. In contrast, defining a corresponding local quantity for the hardness, defined by[9]

$$\eta = \left(\frac{\partial \mu}{\partial N}\right)_{v(r)}, \quad (4)$$

has met essential difficulties, due to the fact that there is no such obvious way to do this as in the case of softness.

A quantitative local hardness concept was first introduced by Berkowitz et al.,[10] who defined the local hardness as

$$\eta(r) = \left(\frac{\delta\mu}{\delta n(r)}\right)_{v(r)}. \tag{5}$$

This local index is then not a local quantity in the sense the local softness is, since it does not integrate to the hardness; consequently, its integral over a given region in the molecule won't necessarily give a regional global hardness. In fact, $\eta(r)$ times the Fukui function is what gives $\eta$ by integration over the whole space,

$$\int \eta(r) f(r) dr = \eta, \tag{6}$$

which emerges via an application of the chain rule. At the same time, Eq.(6) seems to offer some consolation too, yielding the relation

$$\int \eta(r) s(r) dr = 1, \tag{7}$$

via Eq.(3), and utilizing the fact that $S$ is just the inverse of $\eta$,

$$\eta S = 1. \tag{8}$$

Eq.(7) is usually considered as a local version of Eq.(8). However, Eq.(7) cannot be considered as an inverse relation between $\eta(r)$ and $s(r)$ in the chemical viewpoint, since it allows even a possibility that $\eta(r)$ and $s(r)$ are simultaneously small or large in space.

There is, however, another problem with the local hardness defined by Eq.(5). It is not clear how to pinpoint the fixed external potential [$v(r)$] condition on the derivation in Eq.(5). If we consider that the hardness is defined by Eq.(4) as the partial derivative of the chemical potential $\mu[N,v]$ (a function(al) of the electron number $N$ and the external potential) with respect to $N$, Eq.(5) suggests that $v(r)$ *as one of the variables* in $\mu[N,v]$ should be fixed when differentiating with respect to the electron density $\rho(r)$. But this then gives a trivial definition:

$$\eta(\bar{r}) = \left(\frac{\partial\mu[N,v]}{\partial N}\right)_{v(r)} \frac{\delta N}{\delta n(r)} = \eta, \tag{9}$$

*i.e.*, the local hardness equals the global hardness at every point in space.

Utilizing the Euler equation of DFT,

$$\frac{\delta F[n]}{\delta n(r)} + v(r) = \mu, \tag{10}$$

which determines the ground-state density corresponding to a given $v(r)$, $\mu[N,v]$ can be given as

$$\mu[N,v] = \frac{\delta F}{\delta n(r)}[n[N,v]] + v(r) \ . \tag{11}$$

Differentiating this expression with respect to $N$ yields

$$\eta = \left(\frac{\partial \mu[N,v]}{\partial N}\right)_{v(\vec{r})} = \int \frac{\delta^2 F}{\delta n(\vec{r})\delta n(\vec{r}')}\left(\frac{\partial n(\vec{r}')}{\partial N}\right)_{v(\vec{r})} d\vec{r}' = \int \frac{\delta^2 F}{\delta n(\vec{r})\delta n(\vec{r}')} f(\vec{r}') d\vec{r}' \ . \tag{12}$$

Notice the interesting fact that although $\frac{\delta^2 F[n]}{\delta n(r)\delta n(r')}$ is a two-variable function, which is non-constant in either of its variables ($r$ and $r'$), it integrates to a constant when multiplied by the Fukui function. (If it is multiplied by some other function $g(r')$, its integral over $r'$ won't give a constant in $r$ generally.) On the basis of this, it is natural to identify the local hardness given by Eq.(5), and yielding Eq.(9), with

$$\eta(r) = \int \frac{\delta^2 F}{\delta n(r)\delta n(r')} f(r') dr' \ , \tag{13}$$

which gives $\eta$ everywhere. Eq.(13) was proposed by Ghosh,[11] and was discovered to be a constant giving the global hardness at every point (for the ground-state density) by Harbola *et al.*[12]

To propose a different definition for the local hardness than the one yielding the global hardness in every point of space, one may consider the fixed-$v(r)$ constraint in Eq.(5) as a constraint on the differentiation with respect to the density, instead of a simple fixation of the variable $v(r)$ of $\mu[N,v]$. That is, the density domain over which the differentiation is carried out is restricted to the domain of densities that yield the given $v(r)$ – by the first Hohenberg-Kohn theorem,[1] which constitutes a unique $n(r) \to v(r)$ mapping, *i.e.* a $v(r)[n]$ functional. The result will be an ambiguous restricted derivative (see Sec.II of Ref.13 for a general discussion of restricted derivatives), similarly to the case of derivatives restricted to a domain of densities with a given normalization (i.e. particle number), which are determined only up to an arbitrary constant.[1,14] We note that in this case, the notation $\left(\frac{\delta f[...,h,...]}{\delta g(x)}\right)_{h(x)}$ is misleading, since this usually means in physics, and mathematics, a simple fixation of the $h(x)$ variable of the differentiated functional; so the notation

$$\eta(r) = \left. \frac{\delta \mu[N[n], v[n]]}{\delta n(r)} \right|_{v(r)} \qquad (14)$$

may be more appropriate. Above, we also denoted that the dependence of $\mu$ on $n(r)$ is via $N$ and $v(\vec{r})$.

The ambiguity of the local hardness concept was first recognized by Ghosh,[11] and further explored by Harbola et al.,[12] who argued that any $\eta(\vec{r})$ that satisfies Eq.(6) is a good candidate to be taken as the local hardness, and have given an explicit form for this ambiguous $\eta(\vec{r})$:

$$\eta(r) = \int \frac{\delta^2 F}{\delta n(r) \delta n(r')} \lambda(r')[n] dr' , \qquad (15)$$

where $\lambda(\vec{r})$ is an arbitrary function that integrates to 1. Due to the way the second derivative of $F[n]$ delivers the local hardness, and eventually the hardness (via Eq.(6)), it has been termed the hardness kernel. It can be seen that the choice $\lambda(\vec{r}) = f(\vec{r})$ gives back Eq.(13). Another natural choice is $\lambda(\vec{r}) = n(\vec{r})/N$, which yields the original local hardness formula of Berkowitz et al.,[10] who proposed it as a consequence of Eq.(5). Several studies have been made investigating these local hardness concepts and comparing them with each other,[15-18] and numerical evaluations of Eq.(15) have shown that the local hardness can become a useful tool to predict the regioselectivity of chemical reactions[15,16,18,19] and to evaluate the global hardness, using Eq. (6).[20]

Besides the two above definitions for $\eta(r)$, another one has been proposed by Ayers and Parr[17,21] – the unconstrained local hardness,

$$\eta(r) = \frac{\delta \mu[N[n], v[n]]}{\delta n(r)} . \qquad (16)$$

In this case, the fixed-$v(\vec{r})$ constraint on the differentiation with respect to $n(\vec{r})$ is simply dropped, which is a reasonable choice in many aspects. However, this definition brings substantial difficulties when used in practice,[17] due to the explicit appearance of the derivative of $v(r')$ with respect to $n(r)$; therefore, it has been less popular than the older choices of $\eta(r)$. Note that Eq.(16) is embraced by Eq.(14), since for a restricted derivative, a trivial choice is the unrestricted derivative itself (if exists), being valid over the whole functional domain, hence over the restricted domain too.

It is worth pointing out that the hardness kernel itself, too, can be considered as a local hardness, which arises if one fixes the explicit $v(r)$ in Eq.(10) when differentiating $\mu$ with respect to $n(r)$. This case corresponds to the choice $\lambda(r')=\delta(r'-r)$ in Eq.(15), and gives a local quantity that has an extra (parametric) position dependence, showing another strange side of the local hardness concept yielded by Eq.(5). (We have thus a local hardness that can be the global hardness itself and a two space-variable function, too!)

In addition to the above problems, very recently, it has been found[22,23] that the picture of the well-know local descriptors, *i.e.* the local softness and local hardness, is incomplete and the understanding of these reactivity indices must be reconsidered. The largest values of $s(r)$ and $\eta(r)$ do not necessarily correspond to the softest and hardest regions of a molecule, respectively. Rather, $s(r)$ and $\eta(r)$, as defined by Eqs.(3) and (15) (with $\lambda(r)=f(r)$), are pointwise measures of the "local abundance" or "concentration" of the corresponding global quantities. In this framework, $s(r)$ and $\eta(r)$ contain the same potential information and are applicable both to hard and soft systems. In a soft system, $s(r)$ and $\eta(r)$ *both* describe the soft site of a molecule, while in a hard system, $s(r)$ and $\eta(r)$ *both* describe the hard site of the molecule. It therefore seems mandatory to search for alternative definitions for these local descriptors, if the present ones are not adequate reifications of the "chemical" concepts.

As can be seen, the local hardness concept embodied in Eq.(5) poses several deficiencies, which undermines its applicability. In this paper, we will take a new approach to the problem of defining a proper local hardness. Instead of simply replacing $N$ by $n(\bar{r})$ in the definition of hardness (see Eqs.(4) and (5)), which has been proved to be contra productive if one wishes to get a local quantity for the hardness, we will replace $\mu$ by a local chemical potential in Eq.(4), obtaining a local hardness in a similar fashion as a local softness is gained from the softness. In this way, though the ambiguity so characteristic to the local hardness concept cannot be eliminated, we obtain a real local quantity, with desirable features for chemical applications. We will report numerical tests of our new definition of $\eta(r)$ and compare it with the older definitions.

## II. Local hardness through a local chemical potential

The reason for that a local hardness cannot be defined in such a straight way as a local softness is the fact that the chemical potential does not emerge as an integral over space of some "chemical potential density", unlike the electron number, which is the integral of the electron density. However, by utilizing the fact that the chemical potential emerges as the constant term in the *N*-conserving derivative[24,13] of the energy density functional $E_v[n] = F[n] + \int n(\vec{r}) v(\vec{r}) d\vec{r}$ with respect to $n(\vec{r})$, a natural way arises by which the chemical potential is obtained as the space integral of some local quantity. The *N*-conserving derivative of a functional $A[n]$ can be defined as[24,13]

$$\frac{\delta A[n]}{\delta_N n(\vec{r})} = \frac{\delta A[n]}{\delta n(\vec{r})} - \frac{1}{N} \int n(\vec{r}') \frac{\delta A[n]}{\delta n(\vec{r}')} d\vec{r}' \ . \tag{17}$$

This derivative arises from the idea that if two functionals are equal over the domain of $n(\vec{r})$'s of a given norm $N = \int n(\vec{r}) d\vec{r}$, they should have equal derivatives there, too. This constrained derivative concept has found applications in the equations of motion governing the dynamics of two-component thin liquid films,[25] and in the stability analysis of droplet and bubble growth in supercooled vapors and superheated liquids.[26] The appearance of constrained derivatives in (non-variationally derived) physical equations is due to an invariance principle regarding the form of physical equations containing functional derivatives,[27] while their use in the stability analysis of equilibrium is justified by the fact that second-order constrained derivatives properly incorporate all second-order effects due to constraints.[28]

With the use of *N*-conserving differentiation, the Euler equation emerging from the energy minimization principle of DFT for $E_v[n]$ under the constraint of fixation of *N* can be written as[24]

$$\frac{\delta E_v[n]}{\delta_N n(\vec{r})} = 0 \ . \tag{18}$$

This gives

$$\frac{\delta E_v[n]}{\delta n(\vec{r})} = \frac{1}{N} \int n(\vec{r}) \frac{\delta E_v[n]}{\delta n(\vec{r})} d\vec{r} \ , \tag{19}$$

*i.e.*, $\mu$ can be identified as

$$\mu = \frac{1}{N} \int n(\vec{r}) \frac{\delta E_v[n]}{\delta n(\vec{r})} d\vec{r} \ . \tag{20}$$

As $\mu$ is obtained as the integral of a pointwise quantity, this offers a natural way to define a local chemical potential:

$$\mu(\vec{r}) = \frac{n(\vec{r})}{N} \frac{\delta E_v[n]}{\delta n(\vec{r})} \ . \tag{21}$$

Since for ground states, the energy derivative in Eq.(21) is constant in space, and is the chemical potential itself, we obtain

$$\mu(\vec{r}) = \frac{n(\vec{r})}{N}\mu ,  \quad (22)$$

which means that the chemical potential is distributed in space by Eq.(21) according to the distribution of electrons. We emphasize the general validity of the formula Eq.(17), *i.e.* its validity apart from stationary points as well, of which a special case is the minimum situation Eq.(19).

Now, we can define a local hardness by

$$\eta(\vec{r}) = \left(\frac{\partial \mu(\vec{r})[N,v]}{\partial N}\right)_{v(\vec{r})} .  \quad (23)$$

With this definition, the appearance of functional differentiation, and what is more, (ambiguous) restricted functional differentiation, in the local hardness is avoided. It yields the expression

$$\eta(\vec{r}) = \left(f(\vec{r}) - \frac{n(\vec{r})}{N}\right)\frac{\mu}{N} + \frac{n(\vec{r})}{N}\eta .  \quad (24)$$

As can be easily checked, Eq.(24) integrates to $\eta$, with a term integrating to zero correcting the term $\frac{n(\vec{r})}{N}\eta$, which integrates to $\eta$ and is a distribution of the hardness according to the density distribution of electrons. The last term in Eq.(24) is always nonnegative because of the nonnegativity of $\eta$, which is due to the convexity of the energy $E[n,v]$ with respect to the electron number.[1] On the other hand, the other term of Eq.(24) is necessarily negative at some points since it integrates to zero. Its negativity comes from its first term since the density and $-\mu$ is always positive. It may only be negative in those regions of space where the Fukui function is larger than $n(r)/N$.

Eq.(24) is worth rewriting, with the use of Eq.(3), for soft systems, in the form

$$\eta(\vec{r}) + s(\vec{r})\left(-\frac{\mu}{NS}\right) = \frac{n(\vec{r})}{N}\left(\eta - \frac{\mu}{N}\right) ,  \quad (25)$$

where both factors in brackets are positive. (It is important to emphasize that the function which correctly describes the local softness can be taken to be $s(\vec{r}) = Sf(\vec{r})$ only for soft systems.[22,23]) Eq.(25) shows a kind of inverse relationship between local hardness and local softness, in accordance with intuitive expectations, since on the pointwise behavior of local softness (*i.e.* of the derivative of the density with respect to *N*), the pointwise behavior of the density does not have direct effect. If the density is nearly unchanged, for a small $s(\vec{r})$ there will be a large $\eta(\vec{r})$ (in relative terms), and vice versa. To throw more light on the significance of this, integrate Eq.(25) over some volume in space:

$$\Delta\eta + \Delta S\left(-\frac{\mu}{NS}\right) = \frac{\Delta N}{N}\left(\eta - \frac{\mu}{N}\right), \quad (26)$$

where $\Delta$ denotes the part of the global quantity considered that corresponds to the given space volume. Eq.(26) shows that if one divides the electron cloud of a given molecule into (arbitrarily small, or large) pieces consisting of the same number of electrons, the regional hardnesses and softnesses corresponding to these pieces are in an inverse relation over the molecule. In other words, the (local) hardness per electron, $\eta(\vec{r})/\rho(\vec{r})$, can be considered as an inverse of the (local) softness per electron, $s(\vec{r})/\rho(\vec{r})$ (times a constant). It is well-know that the maximum value of the Fukui function $f(r)$ indicates the preferred site for interaction of one system with another.[9] A reactive site implies a soft site, which should be characterized by a large value of $s(r)$ and a small value of $\eta(r)$. Thus, Eq. (24) is a proper quantity for the chemical concept of local hardness with this respect. Of course, Eq.(26) dictates no inverse relation between the local hardnesses and softnesses of two molecules, if considered as individual systems, with separate $N$'s. This is good news; otherwise the local hardness would be an unnecessary quantity – just like the global softness is unnecessary if one has the global hardness (or vice versa), the two being the absolute inverse of each other. It is worth mentioning here the conceptual similarity between $\eta(\vec{r})/\rho(\vec{r})$, or $s(\vec{r})/\rho(\vec{r})$, and the local temperature,[29,30] which is the local (noninteracting) kinetic energy per electron, $t_s(\vec{r})/\rho(\vec{r})$ (times a constant). Note also the importance of the so-called shape function $\frac{n(\vec{r})}{N}$ [14] in Eqs.(22), (24), and (25), which has proved to be a significant quantity in DFT in several aspects.[24,31,32]

Defining a local quantity from an integral expression is of course not without ambiguity. In the case of Eq.(22), as obtained from Eq.(20), this means that an arbitrary function that integrates to zero could be added to $\frac{n(\vec{r})}{N}$. That is, $\frac{n(\vec{r})}{N}$ could be replaced by any other function that integrates to one. This ambiguity actually follows from the $N$-conserving derivative concept too, since in the case of situations where a theory is directly based on the Taylor expansion of its central functional (like in the case of DFT, where its central Euler equation follows from the energy variational principle applied to $E_v[n]$'s Taylor expansion), $\frac{n(\vec{r})}{N}$ can be replaced in Eq.(17) by any other function that integrates to one.[13,28] Consequently, a (mathematically) similar ambiguity arises in $\eta(\vec{r})$'s definition as in the case of the previous local hardness concept of Eq.(5). (There, any function that integrates to zero when multiplied by $f(\vec{r})$ can be added to a given definition of $\eta(\vec{r})$ to obtain another definition.) Conceptually, however, the two cases are different – in the case of Eq.(5), the ambiguity is caused by a fixed-$v(\vec{r})$ restriction on the differentiation with respect to $n(\vec{r})$. In the case of Eq.(23), the two extreme choices of $\eta(\vec{r})$ (the constant $\eta(\vec{r})$ and the two-variable $\eta(\vec{r})$), *e.g.*, are excluded.

The ambiguous function integrating to 1 that may replace $\frac{n(\vec{r})}{N}$ in Eq.(17) can be given as the derivative of a mapping $n_N[n]$ with respect to $N$ (see Appendix of Ref. 28), where $n_N[n]$ integrates to $N$ for any $n(\vec{r})$ and becomes an identity for $n(\vec{r})$'s of norm $N$. That is,

$$\frac{\delta A[n]}{\delta_N n(\vec{r})} = \frac{\delta A[n]}{\delta n(\vec{r})} - \int \frac{\partial n_N(\vec{r}')[n]}{\partial N} \frac{\delta A[n]}{\delta n(\vec{r}')} d\vec{r}' \ . \tag{27}$$

Eq.(17) emerges from $n_N[n] = N \frac{n(\vec{r})}{\int n(\vec{r}')d\vec{r}'}$, *e.g.*, which has the special properties of being homogeneous of degree 0 in $n(\vec{r})$ and leading to the feature of Eq.(17) that it gives back the unconstrained derivative for $N$-independent functionals.[13] Another natural choice is $n_N[n] = n(\vec{r})[N,v[n]]$,[28,33] which yields the Fukui function in the place of $\frac{n(\vec{r})}{N}$ in Eq.(17). With this choice, the local chemical potential arises as

$$\mu(\vec{r}) = f(\vec{r})\mu \ , \tag{28}$$

yielding the local hardness

$$\eta(\vec{r}) = \frac{\partial f(\vec{r})}{\partial N}\mu + f(\vec{r})\eta \ . \tag{29}$$

However, rewriting Eq.(29) in the form $\eta(\vec{r}) + s(\vec{r})\left(\frac{\mu}{S^2}\frac{\partial S}{\partial N} - \frac{\eta}{S}\right) = \frac{\partial s(\vec{r})}{\partial N}\frac{\mu}{S}$ (with $s(\vec{r}) = S f(\vec{r})$), similar to Eq.(25), no such appealing features can be observed as in the case of $\eta(\vec{r})$ given by Eq.(24). It is worth noting here that Eq.(28) can be regained also via

$$\mu = \left(\frac{\partial E_v[n[N,v]]}{\partial N}\right)_{v(\vec{r})} = \int \frac{\delta E_v[n]}{\delta n(\vec{r})}\left(\frac{\partial n(\vec{r})}{\partial N}\right)_{v(\vec{r})} d\vec{r} = \int f(\vec{r})\frac{\delta E_v[n]}{\delta n(\vec{r})} d\vec{r} \ . \tag{30}$$

Finally, we mention that Eq.(24) can be readily generalized for spin-polarized DFT. Eq.(22) will "split" into a local spin-up and spin-down chemical potential,

$$\mu_\uparrow(\bar{r}) = \frac{n_\uparrow(\bar{r})}{N_\uparrow} \mu_\uparrow , \quad (31a)$$

and

$$\mu_\downarrow(\bar{r}) = \frac{n_\downarrow(\bar{r})}{N_\downarrow} \mu_\downarrow , \quad (31b)$$

arising from the SDFT Euler equations

$$\frac{\delta E_{v,B}[n_\uparrow, n_\downarrow]}{\delta_{N_\uparrow} n_\uparrow(\bar{r})} = 0 , \quad (32a)$$

and

$$\frac{\delta E_{v,B}[n_\uparrow, n_\downarrow]}{\delta_{N_\downarrow} n_\downarrow(\bar{r})} = 0 . \quad (32b)$$

The spin-polarized local hardnesses, i.e. local versions of the spin-polarized hardnesses,[34] can then be obtained as

$$\eta_{\uparrow\uparrow}(\bar{r}) = \frac{\partial \mu_\uparrow(\bar{r})[N_\uparrow, N_\downarrow, v, B]}{\partial N_\uparrow} = \left(f_{\uparrow\uparrow}(\bar{r}) - \frac{n_\uparrow(\bar{r})}{N_\uparrow}\right)\frac{\mu_\uparrow}{N_\uparrow} + \frac{n_\uparrow(\bar{r})}{N_\uparrow}\eta_{\uparrow\uparrow} , \quad (33a)$$

$$\eta_{\uparrow\downarrow}(\bar{r}) = \frac{\partial \mu_\uparrow(\bar{r})[N_\uparrow, N_\downarrow, v, B]}{\partial N_\downarrow} = f_{\uparrow\downarrow}(\bar{r})\frac{\mu_\uparrow}{N_\uparrow} + \frac{n_\uparrow(\bar{r})}{N_\uparrow}\eta_{\uparrow\downarrow} , \quad (33b)$$

$$\eta_{\downarrow\uparrow}(\bar{r}) = \frac{\partial \mu_\downarrow(\bar{r})[N_\uparrow, N_\downarrow, v, B]}{\partial N_\uparrow} = f_{\downarrow\uparrow}(\bar{r})\frac{\mu_\downarrow}{N_\downarrow} + \frac{n_\downarrow(\bar{r})}{N_\downarrow}\eta_{\downarrow\uparrow} , \quad (33c)$$

and

$$\eta_{\downarrow\downarrow}(\bar{r}) = \frac{\partial \mu_\downarrow(\bar{r})[N_\uparrow, N_\downarrow, v, B]}{\partial N_\downarrow} = \left(f_{\downarrow\downarrow}(\bar{r}) - \frac{n_\downarrow(\bar{r})}{N_\downarrow}\right)\frac{\mu_\downarrow}{N_\downarrow} + \frac{n_\downarrow(\bar{r})}{N_\downarrow}\eta_{\downarrow\downarrow} . \quad (33d)$$

These local hardnesses have similar inverse relations with the corresponding local softnesses as Eq.(25), as can be seen easily. It has to be emphasized that in order to have derivatives with respect to $N_\uparrow$ and $N_\downarrow$, and previously, with respect to $N$, one must define the energy for noninteger electron numbers. In the case of the ensemble definition of Perdew et al.[35], and its spin-polarized generalization,[36] the derivatives with respect to $N$, and $N_\sigma$'s, or with respect to the corresponding densities, will exhibit discontinuities, leading to a split of the reactivity descriptors into "one-sided" indices (see also Ref. 37).

In the following two sections, we will now compare in detail the local hardness proposed in this section with the older definitions, and make some numerical tests for a few molecular systems.

### III. Computational Details

All calculations were carried out with 6-311+G(2d,2p) basis set[38] at the B3LYP[39] level using the Gaussian 03 package.[40] Three models of local hardness have been evaluated: $\eta(r)[f(r')]$, Eq. (15) with $\lambda(r') = f(r')$, $\eta(r)[\rho(r')/N]$, Eq. (15) with $\lambda(r') = \rho(r')/N$, and the new expression proposed in this work, Eq. (24). (The expression between brackets of $\eta(r)$ indicates the normalized function used to evaluate the local hardness.) Although, some approximations to the universal Hohenberg-Kohn functional and the Fukui function are required. As we have done in our previous articles,[16,22,23] the hardness kernel has been approximated using the second order derivative of the Coulombic-Thomas-Fermi-1/9thWeizsäcker-Dirac-Wigner functional with respect to the density, while the density of the HOMO (highest occupied molecular orbital) can be used as approximation of the Fukui function. Finally, the chemical potential and hardness have been evaluated using the Koopman's approximation, i.e. $\mu = (\varepsilon_{HOMO} + \varepsilon_{LUMO})/2$ and $\eta = \varepsilon_{LUMO} - \varepsilon_{HOMO}$. With these considerations, one obtains the following analytical expressions for $\eta(r)[f(r')]$

$$\eta(r)[n_{HOMO}(r')] = \frac{2}{9} n_{HOMO}(r) n^{-1/3}(\vec{r}) \left[ 5C_K - 2C_X n^{-1/3}(\vec{r}) - 0.0466 \frac{0.458 + 2n^{-1/3}(\vec{r})}{(1 + 0.458 n^{1/3}(\vec{r}))^3} \right] - \frac{1}{36 n(\vec{r})} \nabla \left[ n(\vec{r}) \nabla \left( \frac{n_{HOMO}(r)}{n(\vec{r})} \right) \right] + \int \frac{n_{HOMO}(r')}{|\vec{r} - \vec{r}'|} dr' \quad (34)$$

where $C_K = \frac{3}{10}(3\pi^2)^{2/3}$ and $C_X = \frac{3}{4\pi}(3\pi^2)^{1/3}$ and for $\eta(r)[\rho(r')/N]$

$$\eta(r)[n(r')/N] = \frac{2}{9N} n^{1/3}(\vec{r}) \left[ 5C_K n^{1/3}(\vec{r}) - 2C_X - 0.0466 \frac{0.458 n^{1/3}(\vec{r}) + 2}{(1 + 0.458 n^{1/3}(\vec{r}))^3} \right] + \frac{1}{N} \int \frac{n(\vec{r}')}{|\vec{r} - \vec{r}'|} d\vec{r}' . \quad (35)$$

Finally, the new expression proposed in this work becomes

$$\eta(r) = \left( n_{HOMO}(r) - \frac{n(r)}{N} \right) \frac{\varepsilon_{HOMO} + \varepsilon_{LUMO}}{2N} + \frac{n(r)}{N} (\varepsilon_{LUMO} - \varepsilon_{HOMO}) . \quad (36)$$

In the case of degenerated HOMO orbitals, the Fukui function has been evaluated as the average of the degenerated HOMO orbitals, that is

$$f(\bar{r}) = \frac{1}{d}\sum_{i=1}^{d} n_{HOMO,i}(\bar{r}) \ . \qquad (37)$$

Furthermore, it is possible to evaluate the global hardness of a system, *e.g.* integrating Eq. (36) over the whole space, which results in $\varepsilon_{LUMO} - \varepsilon_{HOMO}$, or combining Eqs. (34) and (35) with Eq. (6), which results in

$$\eta[n_{HOMO}(r), n_{HOMO}(r')] = \int n_{HOMO}(r)\eta(r)[n_{HOMO}(r')]dr \qquad (38)$$

and

$$\eta[n_{HOMO}(r), n(r')/N] = \int n_{HOMO}(r)\eta(r)[n(r')/N]dr \ , \qquad (39)$$

respectively. These integrals have been evaluated numerically using Becke's multicenter integration scheme,[41] which decomposes the integration of a function over the 3D space into a sum of integrations over single-atom components using a weight function, $w_i(r)$, which has the value 1 in the vicinity of its own nucleus, but vanishes in a continuous and well-behaved manner near any other nucleus. The $w_i(r)$ used in this work are the fuzzy Voronoi polyhedra proposed by Becke,[41] taking into account the Bragg–Slater radius[42] and Becke's recipe suggesting to increase the radius of hydrogen to 0.35 Å. Each atom is integrated using Chebyshev's integration for the radial part and Lebedev's quadrature[43] for the angular part. The routine for the Levedev quadrature has been downloaded from Ref. 44. Finally, it is important to remark that the fuzzy Voronoi polyhedra allow the integration over atomic regions, obtaining the condensed atomic hardness, $\eta_i(r)$.[16]

## IV. Results and Discussion

As we have seen, the traditional definitions for the local softness and hardness do not necessarily describe the soft and hard sites of a molecule, respectively. Then, it is necessary to derive new expressions and conditions that the correct local counterparts of the global softness and hardness must fulfill. For the later purpose, we will see that chemical intuition and the concept of global hardness and maximum hardness principle can be very useful tools.

The global hardness represents the resistance of a chemical species to change its electronic configuration; *i.e.* a high value of η means a very stable system to accept and donate electrons. The global hardness is positive and its minimum value is zero. Then, a reasonable first requirement of the correct local hardness is that it must be positive in all positions of the space. Secondly, it must follow the maximum hardness principle (the systems tend to a state of maximum hardness at constant temperature, external potential, and chemical potential). If we consider a local version of this principle, it implies that the most stable regions are also the hardest sites; *i.e.* less reactive and difficult to deform the electron cloud. For instance, the regions in the vicinity of the atomic nuclei must be the hardest regions of a molecule. On the other hand, the softest site of a system is at infinite distance with zero value, where it is unnecessary to provide energy to add or subtract an electron. Another important requisite is that the correct local hardness must show small values at the most reactive places of the molecule, *e.g.* the regions where the molecular frontier orbitals are dominant. According to these conditions, it seems that the local hardness is a function in close proportionality to the electronic density, which establishes a link between $\eta(r)$ and $n(\bar{r})$, and also, a link between conceptual density functional theory and Atoms in Molecules theory.[45] An essential difference between $\eta(r)$ and $n(r)$ of course is that they integrate to the global hardness and the number of electrons, respectively.

In the forthcoming paragraphs, we will check that Eq.(24) follows these requirements. To simplify, we will put $\mu = -\eta/2$, equivalent to considering the electron affinity and $\varepsilon_{LUMO}$ smaller than the ionization potential and $\varepsilon_{HOMO}$, respectively. Eq.(24), then, becomes

$$\eta(r) = -\frac{f(r)}{2N}\eta + \frac{n(r)}{2N^2}\eta + \frac{n(r)}{N}\eta = \frac{\eta}{2N^2}\left[-Nf(r) + [1+2N]n(r)\right]. \tag{40}$$

- At high values of the electronic density, the term which only depends of the Fukui function can be neglected and Eq. (40) results

$$\eta(r) \approx \frac{\eta n(r)}{2N^2}[1+2N], \tag{41}$$

which indicates that the regions close to the atomic nuclei will show maximum values of $\eta(r)$.

- Far away from the molecule, $n(\bar{r})$ and $f(r)$ are zero, and thus, Eq. (40) is also zero.
- The only term of Eqs. (24) and (40), which can be negative, is the term with the Fukui function, considering that the Fukui function is positive (this assumption is generally true, but there also exist some systems, where $f(r)$ can be negative due to orbital relaxation[46]). Then, the reactive sites of the molecule with large values of $f(r)$ will also show small values of $\eta(r)$. If we use the density of the HOMO as approximation of the Fukui function, one can easily see that $\eta(r)$ is a positive function

$$\eta(\vec{r}) = \frac{\eta}{2N^2}\left[-Nn_{HOMO}(\vec{r}) + [1+2N]\sum_{i=1}^{N/2} n_i(\vec{r})\right] \geq 0 \ . \tag{42}$$

To illustrate some of these points, Figures 1, 2, and 3 display the $\eta(r)$ profiles of $Li^+$, $Na^+$, and $K^+$, respectively, using Eqs. (34), (35), and (36). In the case of the $Li^+$ atom, the electronic density is just the density of the HOMO and the three plots of $\eta(r)$ mimic the shape of $n(r)$. Then, the plots of Eqs. (34) and (35) are identical, while in Eq. (36) the first and second terms cancel each other and one obtains $\eta(r) = n_{HOMO}(r)(\varepsilon_{LUMO} - \varepsilon_{HOMO})$. In Figures 2 and 3, one can see that $\eta(r)[n_{HOMO}(r')]$ shows a similar contour than $n_{HOMO}(r)$. On the other hand, $\eta(r)[n(r')/N]$, and the new local hardness proposed in this work, Eq. (36), mimics the contour of $n(r)$. It is worth noting that the shape dependency of Eqs. (34) and (35) with respect to $n_{HOMO}(\vec{r})$ and $n(\vec{r})$, respectively, have already been studied in detail in the literature.[15,16,22,23] The countour of Eq. (36) can be explained analyzing the contributions of its different terms. For instance, at distance of 1 a.u. of the $Na^+$ nucleus the first term $\left(n_{HOMO}(\vec{r})\frac{\varepsilon_{HOMO} + \varepsilon_{LUMO}}{2N}\right)$ contributes with 7.3% to the total value of Eq. (36), while the contribution of the second term $\left(-n(r)\frac{\varepsilon_{HOMO} + \varepsilon_{LUMO}}{2N^2}\right)$ is 8.8% and of opposite sign. This cancellation between these two terms appears along the $\eta(\vec{r})$ profiles of $Na^+$ and $K^+$ atoms, except in the region near to the nucleus, where the term, which depends of $f(r)$, becomes negligible with respect to the other terms.

(Insert Figures 1, 2, and 3 around here)

The Figure 4 shows local hardness profiles parallel to and distance of 0.5 a.u. from the intermolecular axis of the $CO_2$ molecule. The $\eta(r)[n_{HOMO}(r')]$ profile presents two maxima on the oxygens and a minimum on the carbon. In contrast, the local hardnesses of Eqs. (35) and (36) display maxima at the nucleus of the three atoms and two minima (around -0.9 and 0.9 bohrs) near to the two bond critical points (around -0.6 and 0.6 bohrs). From the obtained results from Figures 1-4 and the requirements of $\eta(r)$ established at the beginning of this section, one can easily conclude that $\eta(r)[n(r')/N]$ and the new expression proposed in this work are better reifications of the local counterpart of the global hardness than $\eta(r)[n_{HOMO}(r')]$. In addition, Eq. (36) shows an important advantage with respect to the local hardness as it directly integrates to the global hardness without the requirement to be previously multiplied by the Fukui function.

(Insert Figure 4 around here)

Another interesting example with peculiar behavior between different local hardness profiles is along the $C_6$ axis of benzene, see Figure 5. At the center of the ring, Eqs. (34), (35), and (36) give rise to maximal values, decreasing along the $C_6$ axis. The Coulombic terms of Eqs. (34) and (35) bring a smooth decrease. On the other hand, Eq. (36) only depends of the electronic density [$n_{HOMO}(r)$ is close to zero], and thus, it vanishes around 4 bohrs. In addition, it is worth noting that the term of the Eq. (36), which depends on $1/N^2$, only contributes around 1.5% to the total value of $\eta(r)$ (a common feature to all the systems with large number of electrons).

Figure 6 displays the local hardness profiles along the $C_2$ axis perpendicular to the molecular plane of the ethylene molecule. As one can see, $\eta(r)[n_{HOMO}(r')]$ and $n_{HOMO}(r)$ show maximum values around 1 bohr, where the π–bonding orbital between the two carbons has its most important values. With these results, it seems that we have the contradiction that a reactive site has a large value of local hardness. It is important to remember that $\eta(r)[n_{HOMO}(r')]$ is a function that measures the "local abundance" or "concentration" of $\eta$. Then, the maximum around 1 bohr only indicates that this region will have an important contribution to the value of the global hardness of ethylene. In contrast, the local hardness profiles of Eqs. (35) and (36) show a constant reduction along the $C_2$ axis (more smooth in the case of $\eta(r)[n(r')/N]$). The hardness profile of Eq. (36) for ethylene shows an important difference with respect to the previous systems. It is the first system, where the term of Eq. (36), which depends on $f(r)$, has an important contribution (negative) to the total value of the local hardness, *e.g.* at distances of 1 and 3 bohrs it represents 10% and 30%, respectively. Moreover, one can see the requirement that the regions of the molecule with large values of $f(r)$ (a reactive site) will show a reduction of $\eta(r)$ values.

(Insert Figures 5 and 6 around here)

In a previous article,[22] it has been reported that $\eta(r)[n_{HOMO}(r')]$ of benzocyclobutadiene shows a peculiar behavior. It is a soft molecule, $\eta_{exp}(C_8H_6) = 7.55$ eV,[47] which can be seen as fusion of an aromatic benzene ring and an anti-aromatic cyclobutadiene ring. Using the well-know relationship between aromaticity and hardness,[48] the six-membered ring (Ring1) should be harder than the four-membered ring (Ring2). However three-dimensional contour plots of $\eta(r)[n_{HOMO}(r')]$ and local softness indicate that the largest values of the local hardness and softness are located on the four-membered ring. This molecule is another example that the largest values of $\eta(r)[f(r')]$ do not mean that the four-membered ring is harder in a chemical sense; it only means that the contribution of this ring to the global harndess of benzocyclobutadiene is larger than the contribution of the six-membered ring. To give more insight in this issue, Table 1 contains the global hardness calculated from the three different models of local hardness and the ring hardness. The ring hardness is defined as the average of the condensed atomic hardness of the carbon atoms, which form the ring,

$$\eta_{Ring} = \frac{\sum_{i=1}^{h} \eta_i}{h}, \qquad (43)$$

where $h$ is the number of carbon atoms of the ring. In agreement with our previous results,[22] the ring hardnesses obtained from Eqs. (34) and (35) of the six-membered ring, $\eta_{Ring1}$, are smaller than the four-membered ring, $\eta_{Ring2}$. Eq. (6) explains these results, the Fukui function [in our approach $n_{HOMO}(r)$] is dominant in the four-membered ring and it has a crucial role in the contributions to the global hardness value. In contrast, Eq. (36) directly integrates to the global hardness without the requirement of the Fukui function and as one can see $\eta_{Ring1}$ is slightly larger than $\eta_{Ring2}$, concurring with the compound's chemistry.

(Insert Table 1 around here)

## V. Conclusions

To establish a local measure of chemical hardness is of high significance in conceptual density functional theory. Conventional definitions based on the hardness kernel, though proven to be very useful descriptors to predict regioselectivity of chemical reactions, suffer from some essential drawbacks, which undermine their applicability. First, there is an inconsistency between the local concepts of hardness and softness, as the usual way to originate a local hardness formula does not yield a local quantity that integrates to the global hardness, while the local softness does give the global softness by integration over the whole space of the considered system. In addition, there is a high degree of ambiguity of defining a concrete form of local hardness in the conventional approach of the problem. Moreover, it has been found recently that one of the most-used local hardness expressions, originated from the hardness kernel with the use of the Fukui function, $f(\bar{r})$, contains the same potential information as the local softness concept based on $f(\bar{r})$, instead of being a proper measure of local hardness.

We, therefore, have taken an essentially new approach to find an adequate local counterpart of the chemical hardness concept. Instead of replacing the electron number $N$ by the electron density $n(\bar{r})$, we have replaced the chemical potential $\mu$ by a local chemical potential in the definition of hardness, obtaining a local hardness in a similar fashion as a local softness is gained from the softness. For this, we have utilized the fact that the chemical potential emerges as the additive constant term in the number-conserving functional derivative of the energy density functional. The differentiation of the local chemical potential with respect to $N$ yields a local hardness that integrates to the hardness over the space. An appealing feature of the new local quantity is that it is in a local inverse relation with the Fukui function within a molecule; consequently, it can be an adequate choice for local hardness in the case of soft systems.

We have evaluated three different models of local hardness, namely, the two most-used formulae, $\eta(r)[n_{HOMO}(r')]$ and $\eta(r)[n(r')/N]$, and the new expression proposed in this work, for a few atomic systems and the molecules $CO_2$, $C_2H_4$, benzene, and benzocyclobutadiene. In addition, we have used chemical intuition and the maximum hardness principle to define a set of conditions that an operational local hardness should fulfill. While in the case of $\eta(r)[n_{HOMO}(r')]$, our results confirm previous ones that $\eta(r)[n_{HOMO}(r')]$ is a function that measures the local abundance (or "concentration") of $\eta$, our new expression for $\eta(r)$ and $\eta(r)[n(r')/N]$ have proven to be in good accordance with expectations. Together with its advantageous formal features mentioned above, this indicates that the newly proposed $\eta(r)$ can be a proper local quantity of chemical hardness.

**Acknowledgments:** T.G. acknowledges a visiting postdoctoral fellowship from the Fund for Scientific Research – Flanders (FWO). A grant for T.G. from the Netherlands Fund for Scientific Research is also acknowledged. This research has been supported by the Research

Executive Agency through Grant Agreement n° PERG05-GA-2009-249310 for M.T-S. M.T-S. acknowledges the CSIC for the JAE-DOC contract.**References**

1. R. G. Parr and W. Yang, *Density-Functional Theory of Atoms and Molecules*, Oxford University Press, New York, 1989.
2. H. Chermette, *J. Comput. Chem.*, 1999, **20**, 129; P. Geerlings, F. De Proft and W. Langenaeker, *Chem. Rev.*, 2003, **103**, 1793; J. L. Gázquez, *J. Mex. Chem. Soc.*, 2008, **52**, 3; P. Geerlings and F. De Proft, *Phys. Chem. Chem. Phys.*, 2008, **10**, 3028; S. B. Liu, *Acta Physico-Chimica Sinica*, 2009, **25**, 590; R. K. Roy and S. Saha, *Annu. Rep. Prog. Chem., Sect. C: Phys. Chem.*, 2010, **106**, 118.
3. R. S. Mulliken, *J. Chem. Phys.*, 1934, **2**, 782.
4. R. G. Pearson, *J. Am. Chem. Soc.*, 1963, **85**, 3533; R. G. Pearson, *Science*, 1966, **151**, 172; R. G. Pearson, *Chemical Hardness: Applications from Molecules to Solids*, Wiley-VCH, Oxford, 1997.
5. R. T. Sanderson, *Science*, 1951, **114**, 670.
6. R. G. Pearson, *J. Chem. Educ.*, 1987, **64**, 561; R. G. Parr and P. K. Chattaraj, *J. Am. Chem. Soc.*, 1991, **113**, 1854; R. G. Pearson, *J. Chem. Educ.*, 1999, **76**, 267; M. Torrent-Sucarrat, J. M. Luis, M. Duran and M. Solà, *J. Am. Chem. Soc.*, 2001, **123**, 7951.
7. W. T. Yang and R. G. Parr, *Proc. Natl. Acad. Sci. U.S.A.*, 1985, **82**, 6723.
8. R. G. Parr and W. T. Yang, *J. Am. Chem. Soc.*, 1984, **106**, 4049.
9. R. G. Parr and R. G. Pearson, *J. Am. Chem. Soc.*, 1983, **105**, 7512.
10. M. Berkowitz, S. K. Ghosh and R. G. Parr, *J. Am. Chem. Soc.*, 1985, **107**, 6811.
11. S. K. Ghosh, *Chem. Phys. Lett.*, 1990, **172**, 77.
12. M. K. Harbola, P. K. Chattaraj and R. G. Parr, *Isr. J. Chem.*, 1991, **31**, 395.
13. T. Gál, *J. Math. Chem.*, 2007, **42**, 661.
14. R. G. Parr and L. J. Bartolotti, *J. Phys. Chem.*, 1983, **87**, 2810.
15. P. K. Chattaraj, D. R. Roy, P. Geerlings and M. Torrent-Sucarrat, *Theor. Chem. Acc.*, 2007, **118**, 923.
16. M. Torrent-Sucarrat, P. Salvador, M. Solà and P. Geerlings, *J. Comput. Chem.*, 2008, **29**, 1064.
17. P. W. Ayers and R. G. Parr, *J. Chem. Phys.*, 2008, **128**, 184108.
18. S. Saha and R. K. Roy, *Phys. Chem. Chem. Phys.*, 2008, **10**, 5591.19

**Figure Captions**

**Figure 1:** Local hardness profiles of $Li^+$ using Eqs. (34), (35), and (36) at B3LYP/6-311+G(2d,2p) level. All values are given in a.u.

**Figure 2:** Local hardness profiles of $Na^+$ using Eqs. (34), (35), and (36) at B3LYP/6-311+G(2d,2p) level. All values are given in a.u.

**Figure 3:** Local hardness profiles of $K^+$ using Eqs. (34), (35), and (36) at B3LYP/6-311+G(2d,2p) level. All values are given in a.u.

**Figure 4:** Local hardness profiles parallel to the internuclear axis at a distance of 0.5 a.u. of the $CO_2$ using Eqs. (34), (35), and (36) at B3LYP/6-311+G(2d,2p) level. All values are given in a.u.

**Figure 5:** Local hardness profiles along the $C_6$ axis of benzene using Eqs. (34), (35), and (36) at B3LYP/6-311+G(2d,2p) level, with origin located at the centre of the ring. All values are given in a.u.

**Figure 6:** Local hardness profiles along the $C_2$ axis perpendicular to the molecular plane of ethylene using Eqs. (34), (35), and (36) at B3LYP/6-311+G(2d,2p) level, with origin located at the centre of C=C bond. All values are given in a.u.

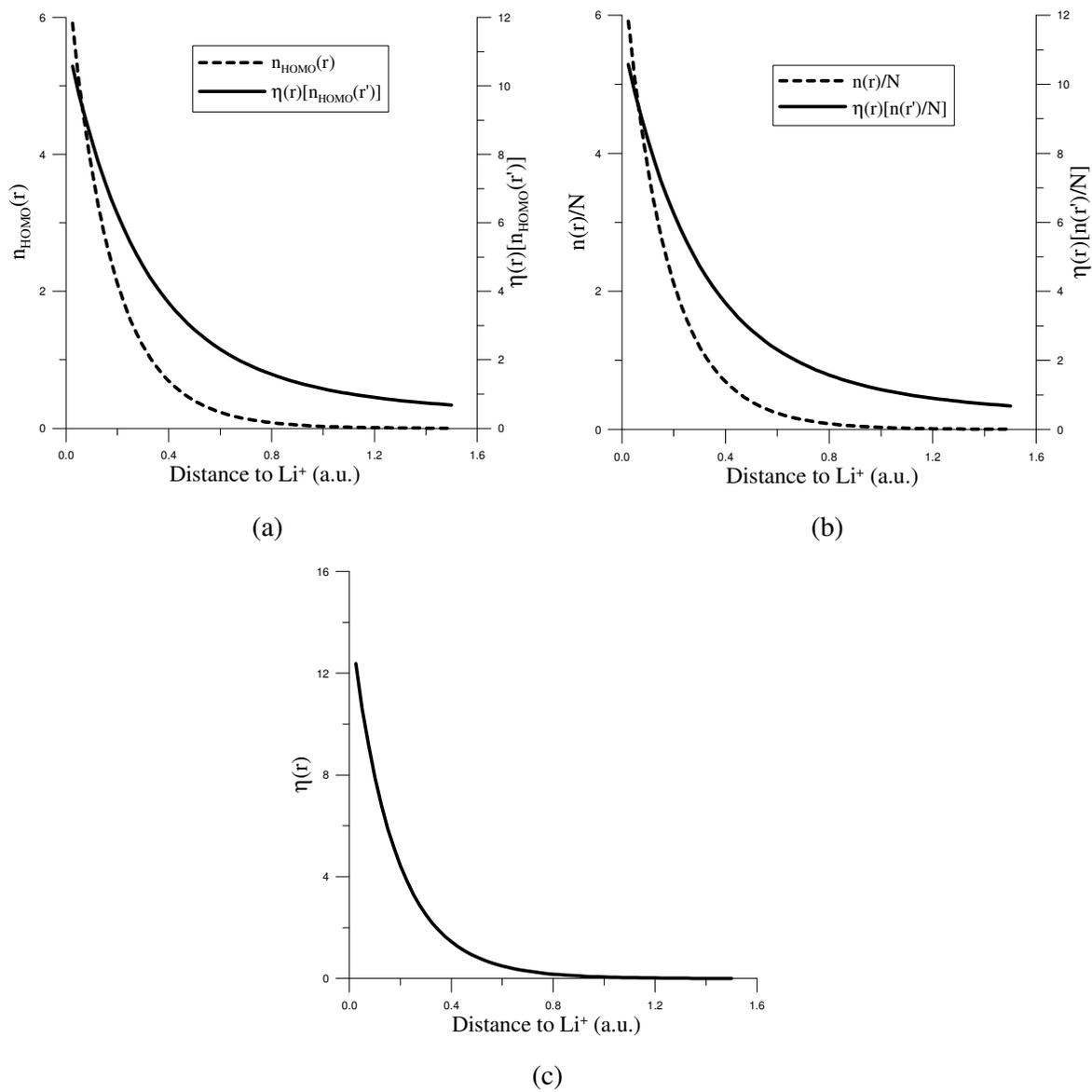

**Figure 1**

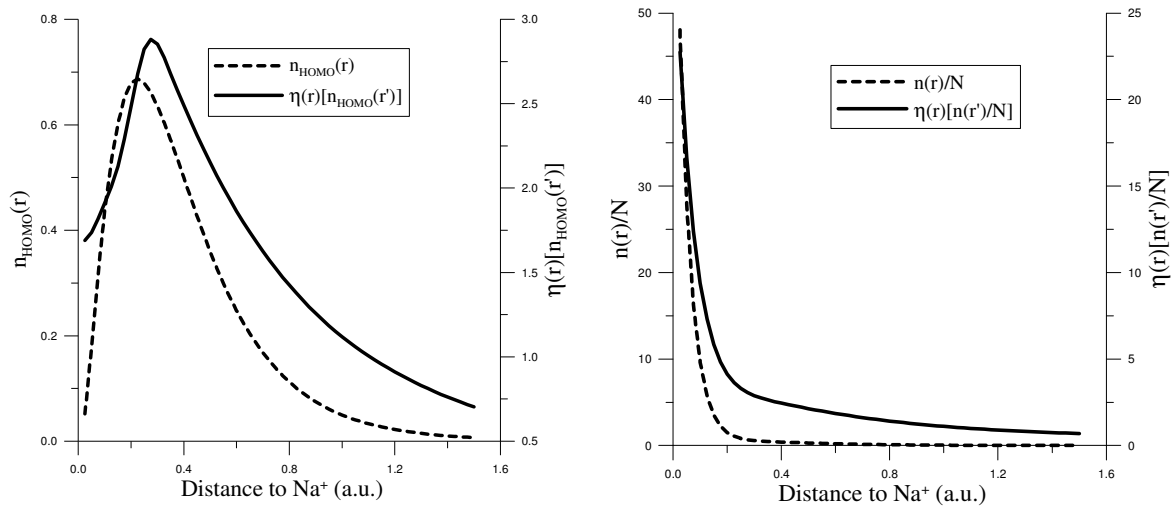

(a)

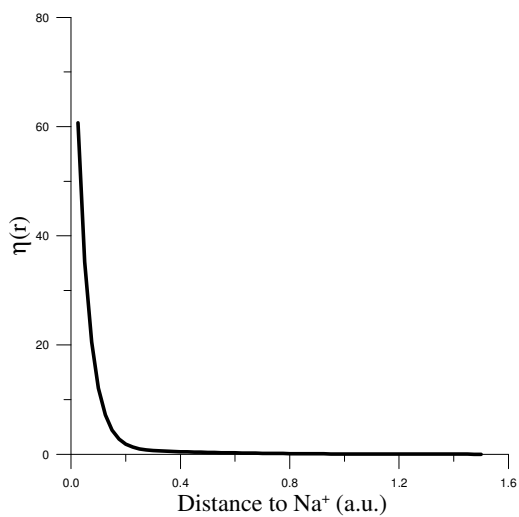

(b)

(c)

# Figure 2

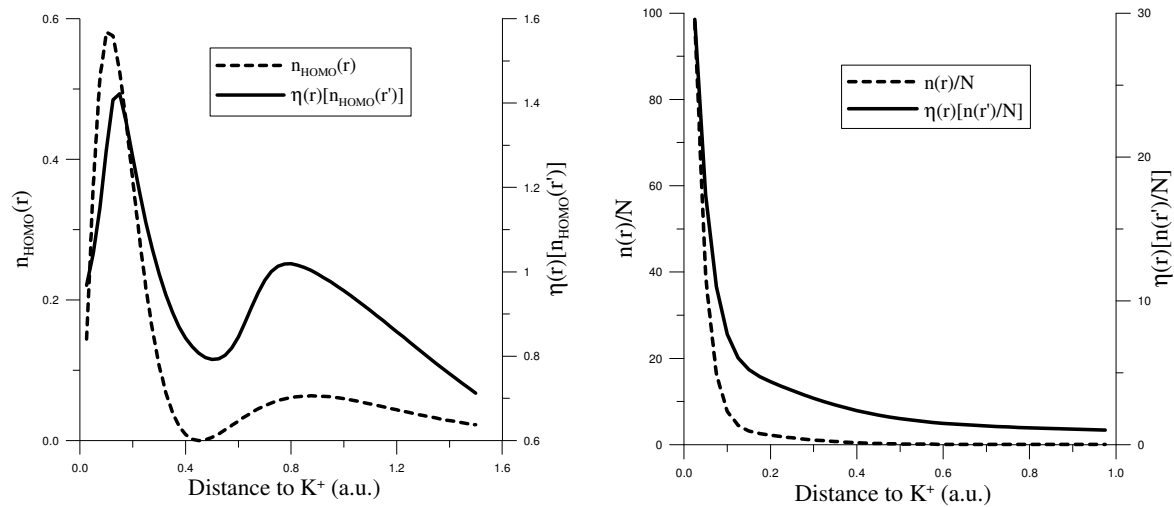

(a)

(b)

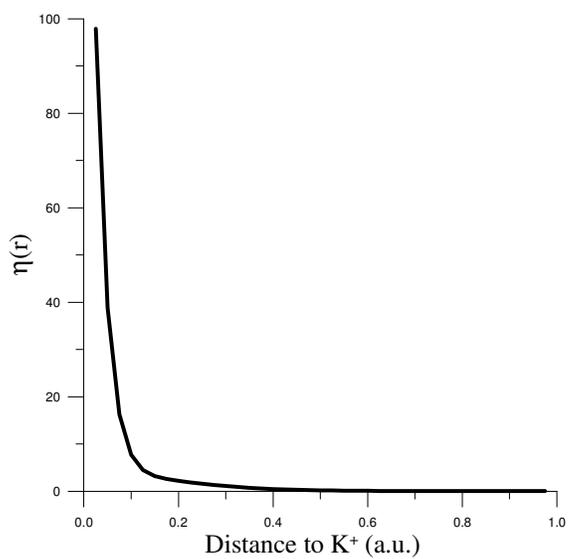

(c)

# Figure 3

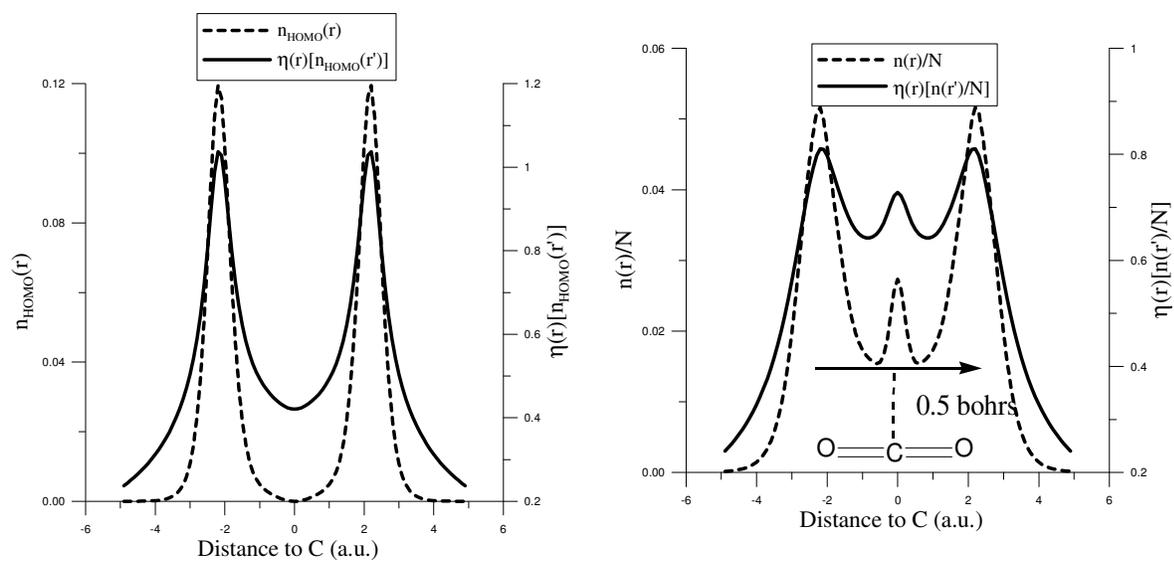

(a)  (b)

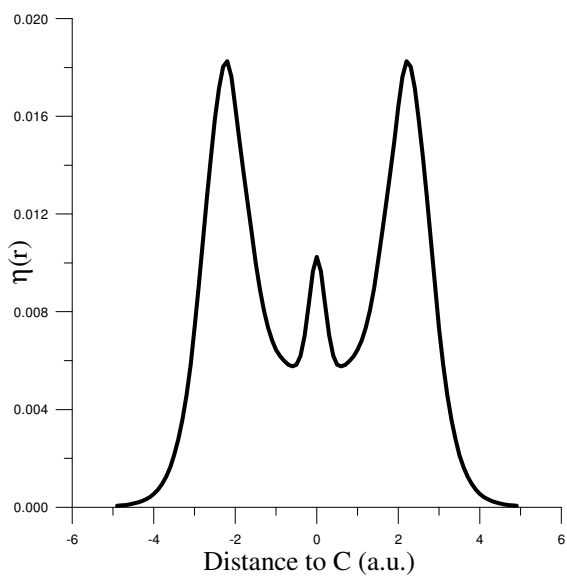

(c)

# Figure 4

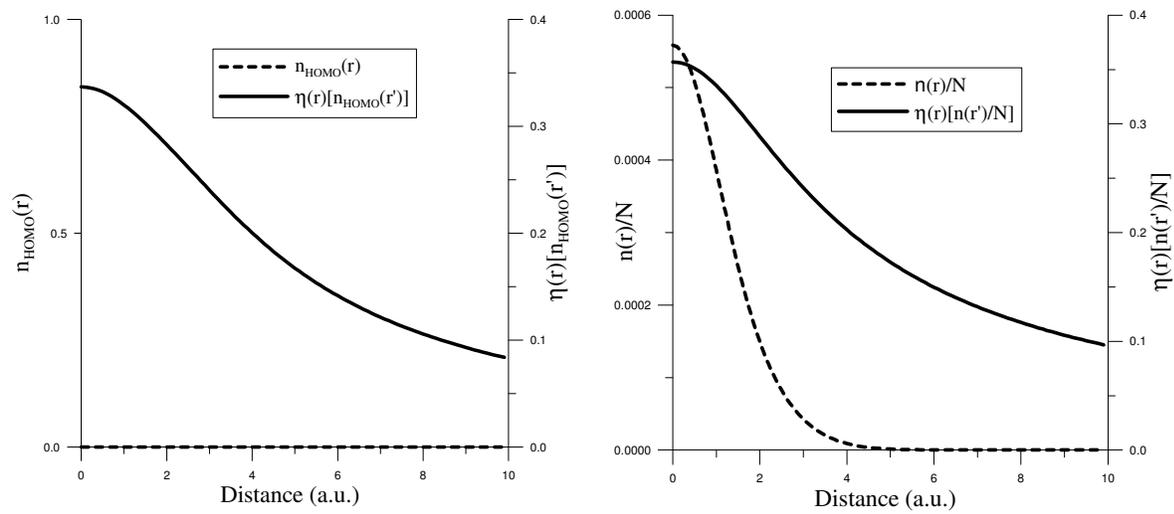

(a)  (b)

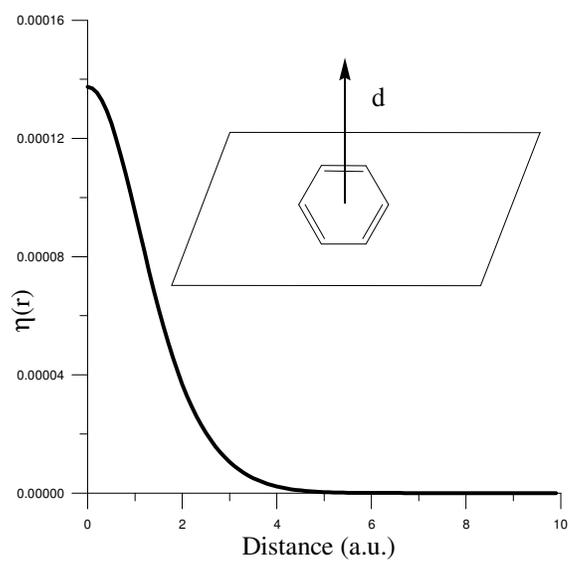

(c)

# Figure 5

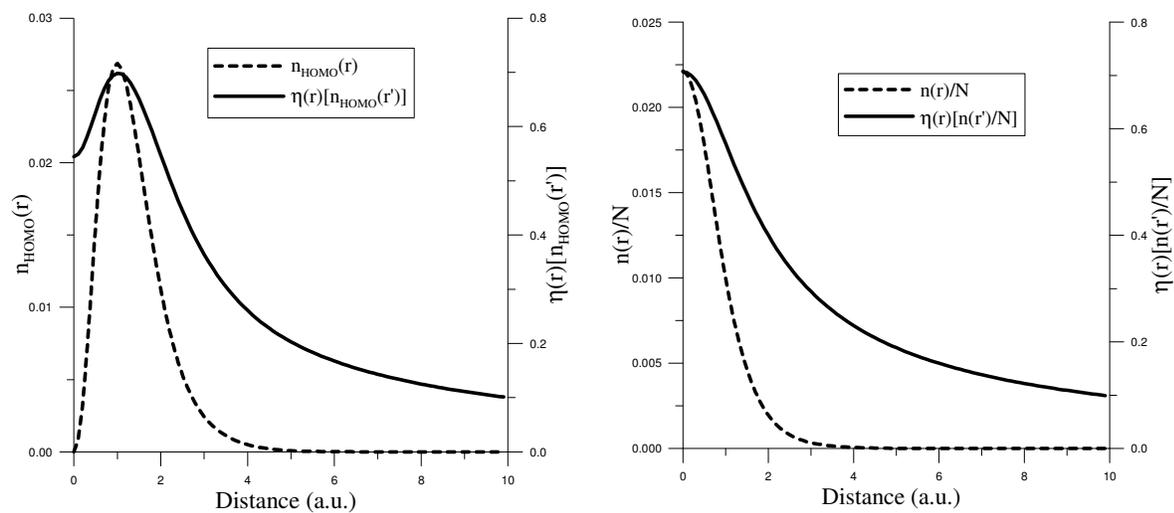

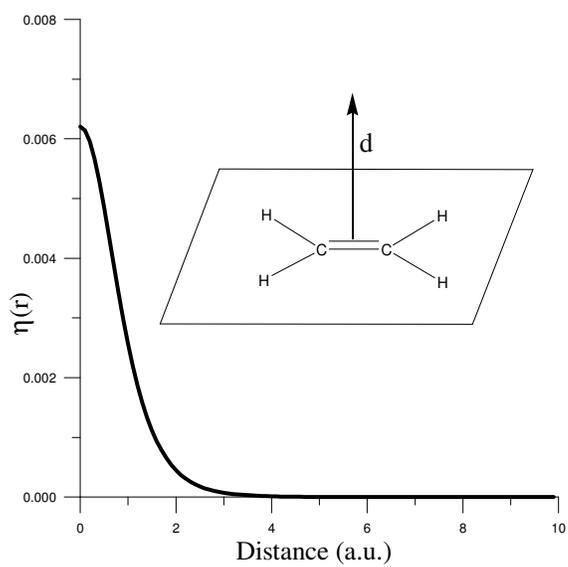

**Figure 6**

**Table 1:** Calculated global and ring hardnesses for the benzocyclobutadine. All values are in eV.

|  | $\eta[n_{HOMO}(r), n_{HOMO}(r')]^a$ | $\eta[n_{HOMO}(r), n(r')/N]^b$ | $\eta^c$ |
|---|---|---|---|
| $\eta_{Ring1}$ | 0.910 | 0.880 | 0.422 |
| $\eta_{Ring1}$ | 1.198 | 0.956 | 0.419 |

[a] Hardness calculated from Eq. (38).
[b] Hardness calculated from Eq. (39).
[c] Hardness calculated from the integration of Eq. (36).